\documentclass[twocolumn,5p,times,sort&compress]{elsarticle}
\usepackage[T1]{fontenc}
\usepackage[utf8]{inputenc}

\usepackage{amsmath,amsfonts,amssymb,bm}
\usepackage[colorlinks,linkcolor=blue,citecolor=blue,urlcolor=blue]{hyperref}

\usepackage[english]{babel}

\usepackage{booktabs}  % table lines

\newcommand{\pT}{p_\text{T}}
\newcommand{\sNN}{\sqrt{s_\text{NN}}}

\newcommand{\wS}{\bm\varpi_\text{S}}
\newcommand{\wT}{\bm\varpi_\text{T}}

\begin{document}

\title{Global spin polarization of multistrange hyperons and feed-down effect\\ in heavy-ion collisions}

\author[KeyLab]{Hui Li}
\ead{lihui\_fd@fudan.edu.cn}

\author[PhyDep]{Xiao-Liang Xia}
\ead{xiaxl@fudan.edu.cn}

\author[KeyLab,PhyDep]{Xu-Guang Huang}
\ead{huangxuguang@fudan.edu.cn}

\author[KeyLab,UCLA]{Huan Zhong Huang}
\ead{huanzhonghuang@fudan.edu.cn}

\address[KeyLab]{Key Laboratory of Nuclear Physics and Ion-beam Application (MOE), Fudan University, Shanghai 200433, China}
\address[PhyDep]{Department of Physics and Center for Field Theory and Particle Physics, Fudan University, Shanghai 200433, China}
\address[UCLA]{Department of Physics and Astronomy, University of California, Los Angeles, CA 90095, USA}

\begin{abstract}
Global spin polarization of hyperons is an important observable to probe the vorticity of the quark-gluon plasma produced in heavy-ion collisions. We calculate the global polarizations of $\Lambda$, $\Xi^-$, and $\Omega^-$ in Au+Au collisions at energies $\sqrt{s_\text{NN}}=$ 7.7--200 GeV based on a multiphase transport model. Our calculations suggest that their primary global polarizations fulfill $P_{\Omega^-}\simeq 5/3P_{\Xi^-}\simeq 5/3P_\Lambda$. We also estimate the feed-down effect of particle decay on the global polarization. With the feed-down effect taken into account, the global polarizations of $\Lambda$ and $\Xi^-$ are clearly separated, and the final global polarizations are in the ordering: $P_{\Omega^-}>P_{\Xi^-}>P_\Lambda$. Such a relation can be tested in experiments, which will provide us more information about the global polarization mechanism.
\end{abstract}

\maketitle

\section{Introduction}\label{sec1}

In noncentral heavy-ion collisions at high energies [$\sNN\sim\mathcal{O}(10)-\mathcal{O}(10^3)$ GeV], an initial angular momentum of the order of $10^4-10^7\hbar$ is generated~\cite{Deng:2016gyh,Jiang:2016woz}. Through the spin-orbit coupling, this angular momentum is partly transformed into the spin of particles~\cite{Liang:2004ph,Voloshin:2004ha}. As a result, particles produced in the collision will have a mean spin polarization, called the \textit{global polarization}, along the initial angular momentum direction.

Experimentally, the spin polarization of hyperons such as $\Lambda$, $\Xi$, and $\Omega$ can be measured by analyzing their parity-violating weak decays~\cite{Lee:1957qs}. The STAR experiment has observed the global polarization of $\Lambda$ ($\overline\Lambda$) hyperon in noncentral Au+Au collisions in the energy region $\sNN=$ 7.7--200 GeV~\cite{STAR:2017ckg,Adam:2018ivw}. Recently, the global polarizations of $\Xi$ and $\Omega$ at $\sNN=$ 200 GeV were also reported~\cite{Adam:2020pti}. It was found that the global polarization of $\Xi$ is slightly larger than that of $\Lambda$, while the global polarization of $\Omega$ seems to be the largest, but the uncertainty is too large to draw a definite conclusion.

Various theoretical approaches (e.g.~Refs.~\cite{Becattini:2007sr,Gao:2007bc,Huang:2011ru,Becattini:2013fla,Fang:2016vpj,Florkowski:2017ruc,Liu:2020flb,Weickgenannt:2020aaf}) have been developed to study the global polarization phenomena in heavy-ion collisions. In most studies, the global polarization is interpreted by coupling between the particle spin and fluid vorticity. Under the assumption of thermal equilibrium of spin degree of freedom, the particle spin polarization can be determined from the thermal vorticity through the following expression~\cite{Becattini:2016gvu}:
\begin{equation}
    \mathbf{P}_\text{H} = \frac{S+1}{3}\left[\frac{E}{m}\wS(x)+\frac{\mathbf{p}}{m}\times\wT(x)-\frac{\mathbf{p}\cdot\wS(x)}{m(E+m)}\mathbf{p}\right].
    \label{PH}
\end{equation}
Here $\mathbf{P}_\text{H}$ is the spin polarization vector of a hyperon defined in its rest frame, which has been normalized to the hyperon spin (i.e.~$\mathbf{P}_\text{H}\equiv\langle\mathbf{S}\rangle/S$ with $\langle\mathbf{S}\rangle$ being the mean spin vector and $S$ the spin quantum number); $E$, $\mathbf{p}$, $m$, and $x$ are the energy, momentum, mass, and space-time coordinate of the hyperon, respectively; and $\wT$ and $\wS$ are components of the thermal vorticity tensor, which is defined as $\varpi_{\mu\nu}=(\partial_\nu\beta_\mu-\partial_\mu\beta_\nu)/2$ where $\beta_\mu=u_\mu/T$ with $u^\mu$ being the fluid velocity and $T$ the temperature. The explicit expressions of $\wT$ and $\wS$ are
\begin{align}
    \wT & = (\varpi_{tx},\varpi_{ty},\varpi_{tz}) = \frac{1}{2}(\nabla\beta_t+\partial_t\bm\beta), \label{WT} \\
    \wS & = (\varpi_{yz},\varpi_{zx},\varpi_{xy}) = \frac{1}{2}\nabla\times\bm\beta.    \label{WS}
\end{align}
Based on Eqs.~(\ref{PH})--(\ref{WS}), the global $\Lambda$ polarization has been calculated by using relativistic hydrodynamics~\cite{Karpenko:2016jyx,Xie:2017upb,Fu:2020oxj} and coarse-grained transport models~\cite{Li:2017slc,Shi:2017wpk,Xia:2018tes,Wei:2018zfb,Vitiuk:2019rfv}. The calculations successfully described the experimental data~\cite{STAR:2017ckg,Adam:2018ivw} of the collision energy dependence of the global $\Lambda$ polarization. Nevertheless, how spin of particles reaches the thermal equilibrium in a strongly interacting quark-gluon plasma so that Eq.~(\ref{PH}) can apply is still an open question which is under intense investigation~\cite{Hattori:2019lfp,Bhadury:2020puc,Li:2019qkf,Ayala:2019iin,Ayala:2020ndx,Kapusta:2019ktm,Kapusta:2019sad,Wang:2021qnt}. Therefore, more theoretical and experimental results of the global polarization are essential to verify the current vorticity interpretation.

In this paper, we calculate and compare the global polarizations of $\Lambda$, $\Xi^-$, and $\Omega^-$ hyperons. These hyperons are different in spin, mass, and constituent quark flavor. Moreover, as shown in Sect.~\ref{sec3}, they also receive different contributions from decay of heavier particles. Therefore, it is natural to expect that the three hyperons have different global polarizations, but if Eq.~(\ref{PH}) is valid, their global polarizations should satisfy certain relations. In this paper, we focus on revealing these relations, which are expected to be verified in experiments.

This paper is organized as follows. We first calculate the global polarizations of the three hyperons using Eqs.~(\ref{PH})--(\ref{WS}). The calculation method is briefly described in Sect.~\ref{sec2.1}, and the results and discussion are presented in Sect.~\ref{sec2.2}. We note that the global polarizations calculated by Eq.~(\ref{PH}) are those of particles formed by hadronization, so they are called the primary polarizations. In experiments, however, a considerable amount of final particles are produced by decay of heavier particles. Therefore, we also investigate the feed-down effect of the decays on the global polarization in Sect.~\ref{sec3}. Finally, conclusions are drawn in Sect.~\ref{sec4}.

\section{Primary global polarization}\label{sec2}

\subsection{Model setup}\label{sec2.1}

In this section, we use the string-melting version of a multi-phase transport (AMPT) model~\cite{Lin:2004en,Lin:2014tya} to calculate the thermal vorticity and the primary global polarization. This model, together with the calculation method described below, has been widely used in previous studies of the $\Lambda$ polarization~\cite{Li:2017slc,Shi:2017wpk,Xia:2018tes,Wei:2018zfb}. It produced a reasonable collision energy dependence of the global $\Lambda$ polarization, which is in agreement with the experimental data; see Ref.~\cite{Huang:2020dtn} for a review.

Here, we briefly describe the method to calculate the thermal vorticity and the primary global polarization with the AMPT model. As a transport model, the AMPT explicitly tracks each particle's position and momentum during the evolution of the collision system. Therefore, a natural way to calculate the thermal vorticity in this model is by using the coarse-grained method. By splitting the whole space-time volume into grid cells, the energy-momentum tensor $T^{\mu\nu}$ can be calculated by an event average of $p^\mu p^\nu/p^0$ of all particles in each cell:
\begin{equation}
    T^{\mu\nu}(\tau,x,y,\eta_s)=\frac{1}{\Delta V}\left\langle \sum_i\frac{p_i^\mu p_i^\nu}{p_i^0}\right\rangle,
\end{equation}
where $\Delta V$ is the space volume of the cell, the index $i$ labels the $i$-th particle inside the cell, and the angle brackets represent the event average, which is taken to cancel the random thermal motion of particles, retaining only the collective motion. From the obtained $T^{\mu\nu}$, the fluid velocity $u^\mu$ and the energy density $\varepsilon$ can be computed by numerically solving the eigenvalue equation $T^{\mu\nu} u_\nu=\varepsilon u^\mu$ with a normalization condition $u^\mu u_\mu=1$. The energy density $\varepsilon$ is used to determine the temperature $T$ through the lattice equation of state~\cite{Borsanyi:2013bia}. With the obtained velocity and temperature, the thermal vorticity field is calculated by using Eqs.~(\ref{WT})--(\ref{WS}).

To calculate the primary global polarization, we further need to know the phase-space distribution of hyperons when they are formed by hadronization. In the AMPT model, the hadronization is implemented by coalescence of quarks. In this process, the space-time coordinates and energy-momentum vectors of all hadrons are recorded. This allows us to directly calculate the spin polarization vector of every hadron using Eq.~(\ref{PH}) with the hadron's energy and momentum, and with the thermal vorticity value at the hadron's formation location. Finally, by projecting the spin polarization vector along the direction of the initial angular momentum of the collision system, and by averaging it in a specific kinematic region, the global polarization for a given hyperon is obtained as
\begin{equation}
    P_\text{H} = \langle \mathbf{P}_\text{H} \cdot \hat{\mathbf{J}} \rangle,
\end{equation}
where $\hat{\mathbf{J}}$ stands for the unit vector along the initial angular momentum direction.

We run simulations for Au+Au collisions at energies $\sNN=$ 7.7, 11.5, 14.5, 19.6, 27, 39, 62.4, and 200 GeV. At each collision energy, $10^6$ events are generated to extract the thermal vorticity. In the coarse-grained procedure, we work in the Milne coordinates $x^\mu=(\tau,x,y,\eta_s)$. The space-time volume is divided into 60 time steps with the interval $\Delta\tau=0.5$ fm/c, $80\times 80$ spatial grids with $\Delta x=\Delta y=0.5$ fm, and 40 space-time rapidity steps with $\Delta\eta_s=0.25$. With the extracted thermal vorticity field, we calculate the primary global polarizations of $\Lambda$, $\Xi^-$, and $\Omega^-$ using the aforementioned method. In the calculation, the global polarizations are averaged in the kinematic region of rapidity $|y|<1$ and transverse momentum $\pT>0.5$ GeV/c, which approximately match the STAR experimental acceptance~\cite{STAR:2017ckg,Adam:2018ivw,Adam:2020pti}.

\subsection{Results and discussion}\label{sec2.2}

\begin{figure}
    \centering
    \includegraphics[width=1\columnwidth]{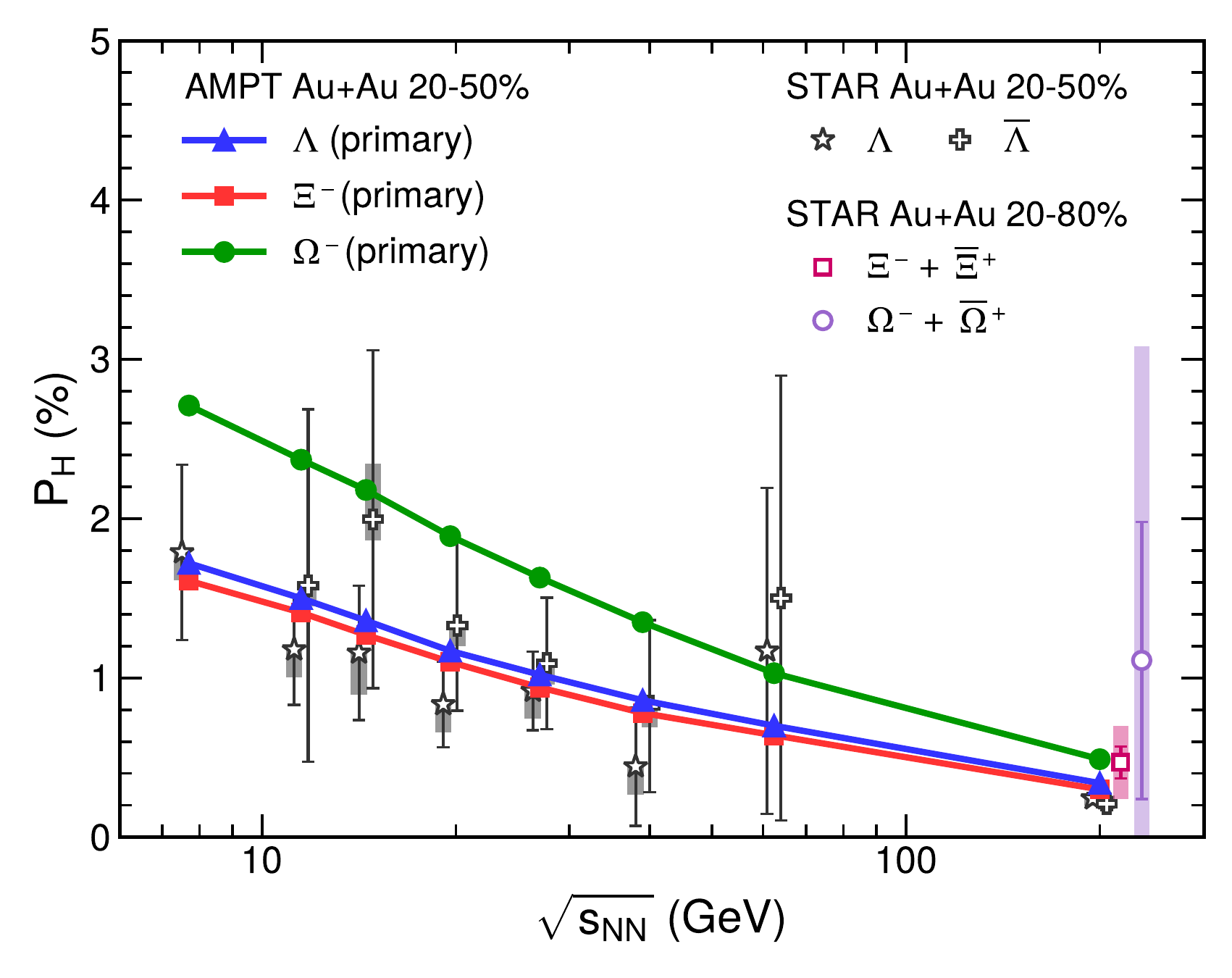}
    \caption{(Color online) Collision energy dependence of the primary global polarizations of $\Lambda$, $\Xi^-$, and $\Omega^-$ hyperons in 20--50\% central Au+Au collisions at energies $\sNN=$ 7.7--200 GeV. The STAR experimental data~\cite{STAR:2017ckg,Adam:2018ivw} for global $\Lambda$ and $\overline\Lambda$ polarization in the same energy and centrality range, as well as the recent data~\cite{Adam:2020pti} for global $\Xi$ and $\Omega$ polarization at 200 GeV for 20--80\% centrality, are also shown.}
    \label{fig-PH}
\end{figure}

Figure~\ref{fig-PH} shows the collision energy dependence of the primary global polarizations of $\Lambda$, $\Xi^-$, and $\Omega^-$ hyperons in 20--50\% central Au+Au collisions in the energy region $\sNN=$ 7.7--200 GeV. The primary global polarizations of the three hyperons all decrease with the increase of collision energy. The decreasing global $\Lambda$ polarization has been observed both theoretically~\cite{Karpenko:2016jyx,Xie:2017upb,Fu:2020oxj,Li:2017slc,Shi:2017wpk,Xia:2018tes,Wei:2018zfb,Vitiuk:2019rfv} and experimentally~\cite{STAR:2017ckg,Adam:2018ivw} in previous studies. In the vorticity interpretation of the global polarization, the decreasing behavior can be understood from the fact that a smaller vorticity is generated in the mid-rapidity region at higher collision energies~\cite{Deng:2016gyh,Jiang:2016woz}.

Let us focus on the hyperon-species dependence of the global polarization. From Fig.~\ref{fig-PH}, the primary global polarization of $\Omega^-$ is larger than those of $\Lambda$ and $\Xi^-$, and the primary global polarizations of $\Lambda$ and $\Xi^-$ are very close to each other. This finding is consistent with the recent hydrodynamic calculation preformed in Ref.~\cite{Fu:2020oxj}. The difference between the primary global polarization of $\Omega^-$ and those of $\Lambda$ and $\Xi^-$ mainly arises from their spin numbers, which are $S_{\Omega^-}=3/2$ and $S_\Lambda=S_{\Xi^-}=1/2$. According to Eq.~(\ref{PH}), the hyperon polarization is proportional to $S+1$. Considering this effect only, the primary global polarization of $\Omega^-$ should be $5/3$ times of those of $\Lambda$ and $\Xi^-$. This relation is approximately satisfied  in Fig.~\ref{fig-PH}, i.e.~$P_{\Omega^-}\simeq 5/3P_{\Xi^-}\simeq 5/3P_\Lambda$.

Besides the spin number, $\Lambda$, $\Xi^-$, and $\Omega^-$ also differ in mass and constituent quark flavor, which in general can lead to different distributions of their velocity and formation location. Note that $\mathbf{P}_\text{H}$ in Eq.~(\ref{PH}) depends on the four velocity $(E/m,\mathbf{p}/m)$ and the formation location $x$ of the hyperon. As a result, the global polarizations of primary $\Lambda$ and $\Xi^-$ are not necessarily the same, though they have the same spin. However, as demonstrated in Fig.~\ref{fig-PH}, the difference between the primary global polarizations of $\Lambda$ and $\Xi^-$ is very small. By contrast, the feed-down effect as shown in Sect.~\ref{sec3} can cause a much larger difference.

\begin{figure}
    \centering
    \includegraphics[width=1\columnwidth]{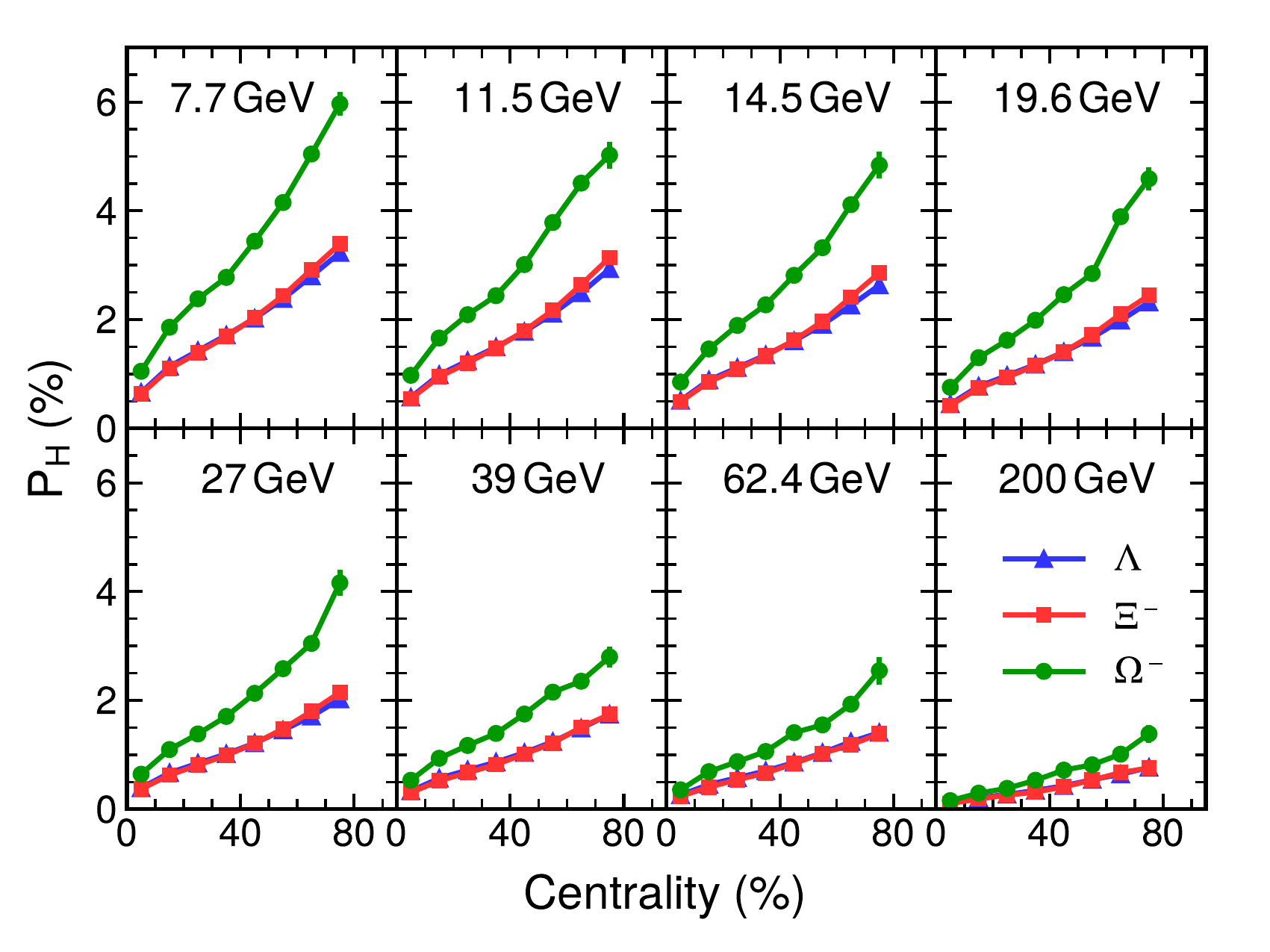}
    \caption{(Color online) Centrality dependence of the primary global polarizations of $\Lambda$, $\Xi^-$, and $\Omega^-$ in Au+Au collisions at $\sNN=$ 7.7--200 GeV.}
    \label{fig-cen}
\end{figure}

In Fig.~\ref{fig-cen}, we present the centrality dependence of the primary global polarizations of $\Lambda$, $\Xi^-$, and $\Omega^-$ at energies $\sNN=$ 7.7--200 GeV. At each collision energy, we find again that the primary global polarization of $\Omega^-$ is larger than those of $\Lambda$ and $\Xi^-$ and the latter two are very close. We also find that the global polarizations increase with the increasing centrality, which is a natural expectation of the vorticity interpretation of the global polarization.

The results presented in Figs.~\ref{fig-PH} and~\ref{fig-cen} include only the spin polarization induced by the thermal vorticity tensor [see Eq.~(\ref{PH})]. Recently, it was found that the shear tensor also contributes to the spin polarization~\cite{Liu:2021uhn,Becattini:2021suc}. Hydrodynamic simulations~\cite{Fu:2021pok,Becattini:2021iol,Yi:2021ryh} show that the shear tensor essentially affects the local spin polarization (the polarization as a function of momentum). However, to the global spin polarization, in which the momentum is integrated, the contribution from the shear tensor is negligible~\cite{Ryu:2021lnx,Liu:2021nyg}. In this study, we have also calculated the global polarization induced by the shear tensor using the AMPT model, and found its contribution is almost zero.

In the calculations of the global polarization, we have adopted a scenario that hyperon's spin polarization directly responds to the thermal vorticity at hadronization by Eq.~(\ref{PH}). We note that another scenario, in which quarks respond to the thermal vorticity and hadrons inherit the quark polarization, was used in the recent studies~\cite{Xia:2020tyd,Fu:2021pok,Yi:2021ryh}. In this scenario, the quark model indicates that the hyperon polarizations are $P_\Lambda= P_s$, $P_{\Xi^-}\approx (4P_s-P_d)/3$, and $P_{\Omega^-}\approx 5 P_s/3$~\cite{Liang:2004ph,Yang:2017sdk}, where $P_s$ and $P_d$ are the polarizations of $s$ and $d$ quarks, respectively. In principle, quarks of different flavors may have different global polarizations, so $P_\Lambda$ and $P_{\Xi^-}$ may be different. However, if $P_s$ and $P_d$ are close in the most likely case, the quark-polarization scenario and our hadron-polarization scenario will have the same prediction of $P_\Lambda:P_{\Xi^-}:P_{\Omega^-}\approx 1:1:5/3$. Precision measurements of $P_\Lambda$ and $P_{\Xi^-}$ could potentially determine the deviation between strange and down quark polarizations.

\section{Feed-down effect}\label{sec3}

The global polarizations studied in Sect.~\ref{sec2} are those of the primary hyperons which are produced by hadronization. However, in realistic heavy-ion collisions, a considerable amount of final particles are produced by decay of heavier particles. In the vorticity interpretation of the global polarization, the parent particles also possess the global polarization. When they decay, part of their global polarizations are transferred to the decay products. As a result, the global polarization of inclusive hyperons in the final state could be different from that of primary ones. The feed-down effect on the global $\Lambda$ polarization has been estimated in Refs.~\cite{Becattini:2016gvu,Karpenko:2016jyx,Li:2017slc}, and the feed-down effect on the azimuthal-angle dependent $\Lambda$ polarization has been studied in Refs.~\cite{Xia:2019fjf,Becattini:2019ntv}.

In this section, we estimate the feed-down effect on the global $\Xi^-$ polarization in Au+Au collisions at 7.7--200 GeV. We also recalculate the feed-down effect on the global $\Lambda$ polarization. We note that very rare particles can decay to $\Omega^-$. Therefore, the feed-down effect on the global $\Omega^-$ polarization is very small and is neglected in this study.

\begin{table}[t]
    \caption{The polarization transfer factor $C$ [defined in Eq.~(\ref{transfer})] of several important decays}
    \label{table-C}
    \begin{center}
        \begin{tabular}{lr}
            \toprule
            Decay                                            & $C$                         \\
            \midrule
            Strong decay $1/2^+\to 1/2^+\,0^-$               & $-1/3$                      \\
            Strong decay $1/2^-\to 1/2^+\,0^-$               & $1$                         \\
            Strong decay $3/2^+\to 1/2^+\,0^-$               & $1$                         \\
            Strong decay $3/2^-\to 1/2^+\,0^-$               & $-3/5$                      \\
            Electromagnetic decay $\Sigma^0\to\Lambda\gamma$ & $-1/3$                      \\
            Weak decay $\Xi^-\to\Lambda\pi^-$                & $(1+2\gamma)/3\approx0.944$ \\
            Weak decay $\Xi^0\to\Lambda\pi^0$                & $(1+2\gamma)/3\approx0.915$ \\
            \bottomrule
        \end{tabular}
    \end{center}
\end{table}

How spin polarization transfers from the parent particle $P$ to the decay product $D$ in a two-body decay $P\to D+X$ has been studied in Refs.~\cite{Becattini:2016gvu,Xia:2019fjf,Becattini:2019ntv}. In general, the polarization of $D$ depends not only on the polarization of $P$, but also on the momentum direction of $D$ in the decay rest frame (see Refs.~\cite{Xia:2019fjf,Becattini:2019ntv} for detailed spin transfer rules). However, if we are only interested in the global polarization, in which the momentum of $D$ is integrated, the global polarizations of $P$ and $D$ satisfy a simple linear relation:
\begin{equation}
    \mathbf{P}_D=C\mathbf{P}_P.
    \label{transfer}
\end{equation}
In this equation, the global polarizations $\mathbf{P}_D$ and $\mathbf{P}_P$ are defined in the rest frames of $D$ and $P$, respectively; and $C$ is the polarization transfer factor. The values of $C$ for the decay channels relevant in our calculations are listed in Table~\ref{table-C}. These values can be found in Refs.~\cite{Becattini:2016gvu,Xia:2019fjf,Becattini:2019ntv}
\footnote{In Ref.~\cite{Xia:2019fjf}, the factor $C$ is defined as the polarization transfer factor $C=\mathbf{P}_D/\mathbf{P}_P$; while in Refs.~\cite{Becattini:2016gvu,Becattini:2019ntv}, the factor $C$ is defined as the spin transfer factor $C=\mathbf{S}_D/\mathbf{S}_P$. Different values of $C$ are reported in Ref.~\cite{Xia:2019fjf} and in Refs.~\cite{Becattini:2016gvu,Becattini:2019ntv}, but they are consistent with each other because $\mathbf{P}=\mathbf{S}/S$. In this paper, we use the same definition of $C$ as in Ref.~\cite{Xia:2019fjf}.}.
For the strong and electromagnetic decays listed in the table, $C$ are pure numbers, which are completely determined by the spin and parity of the parent and product particles of the decay. However, for the weak decay $\Xi\to\Lambda\pi$, $C$ depends on a decay parameter $\gamma$, which is defined as
\begin{equation}
    \gamma = \frac{|A_\text{s}|^2-|A_\text{p}|^2}{|A_\text{s}|^2+|A_\text{p}|^2},
    \label{gamma}
\end{equation}
where $A_\text{s}$ and $A_\text{p}$ are known as the partial-wave amplitudes for decay via s and p wave, respectively. Using the experimentally measured values of $\gamma$ available in the Particle Data Group (PDG) book~\cite{Zyla:2020zbs}, we obtain $C\approx 0.944$ for $\Xi^-\to\Lambda\pi^-$ and $C\approx 0.915$ for $\Xi^0\to\Lambda\pi^0$.

Now we study the feed-down effect on the global $\Xi^-$ polarization. The decay channel to $\Xi^-$ is mainly the strong decay $\Xi(1530)\to\Xi^-\pi$. [Here and the same as below, $\Xi(1530)^0\to\Xi^-\pi^+$ and $\Xi(1530)^-\to\Xi^-\pi^0$ are combined]. In this decay, the spin and parity is $3/2^+\to 1/2^+\,0^-$. As shown in Table~\ref{table-C} the polarization transfer factor $C$ is one, which means that the decay product $\Xi^-$ can fully inherit the global polarization of $\Xi(1530)$. Considering that $\Xi(1530)$ is spin 3/2, the primary global polarization of $\Xi(1530)$ should be approximately 5/3 times of that of $\Xi^-$ as implied by Eq.~(\ref{PH}). Putting all these together, the global polarization of $\Xi^-$ with the feed-down contribution is related to its primary global polarization, as
\begin{align}
    P_{\Xi^-}(\text{primary+feed-down})=\quad                                          & \nonumber                  \\
    \frac{N_{\Xi^-}+\frac{5}{3}N_{\Xi(1530)\to\Xi^-}}{N_{\Xi^-}+N_{\Xi(1530)\to\Xi^-}} & P_{\Xi^-}(\text{primary}),
    \label{P-corr}
\end{align}
where $N_{\Xi^-}$ is the particle number of primary $\Xi^-$ and $N_{\Xi(1530)\to\Xi^-}$ is the number of $\Xi^-$ from the decay of $\Xi(1530)$. From Eq.~(\ref{P-corr}), the global polarization of inclusive $\Xi^-$ should be larger than that of the primary $\Xi^-$.

\begin{table}[t]
    \caption{The feed-down contribution factor $\delta_\text{H}$ [defined by Eq.~(\ref{FD-effect})] on the global polarizations of $\Lambda$ and $\Xi^-$ at different collision energies}
    \label{table-feed-down}
    \begin{center}
        \begin{tabular}{lcc}
            \toprule
            $\sNN$\,(GeV) & $\delta_\Lambda$ & $\delta_{\Xi^-}$ \\
            \midrule
            7.7           & $-13.2\%$        & $+24.2\%$        \\
            11.5          & $-11.6\%$        & $+25.8\%$        \\
            19.6          & $-10.3\%$        & $+26.4\%$        \\
            27            & $-9.7\%$         & $+26.7\%$        \\
            39            & $-9.3\%$         & $+26.7\%$        \\
            62.4          & $-9.2\%$         & $+26.4\%$        \\
            200           & $-8.3\%$         & $+27.1\%$        \\
            \bottomrule
        \end{tabular}
    \end{center}
\end{table}

To quantitatively estimate the above feed-down contribution, we need to calculate the primary yields of $\Xi^-$ and $\Xi(1530)$. We note that the quark coalescence in the AMPT model does not produce baryon excited states $\Xi(1530)$, $\Sigma(1385)$, etc.~\cite{Lin:2004en}. Therefore, we use a thermal statistical model, THERMUS~\cite{Wheaton:2004qb}, to estimate the relative yield of primary particles, which has been tuned to successfully describe the particle yields in relativistic heavy-ion collisions. The main parameters of this model are the chemical freeze-out temperature, the chemical potentials for baryon and strangeness, and the strangeness suppression factor. In our calculation, we use the same set of the parameters provided in Ref.~\cite{Adamczyk:2017iwn}, which was tuned to reproduce the experimental data at $\sNN=$ 7.7--200 GeV. We also update the particle list of the THERMUS model according to the latest updated PDG book of the year 2020~\cite{Zyla:2020zbs}. According to the calculation, the ratio of $N_{\Xi^-}$ to $N_{\Xi(1530)\to\Xi^-}$ is about 59 : 41 in 20--50\% central Au+Au collisions at 200 GeV. With the yield ratio, we can calculate feed-down contribution factor, defined as
\begin{equation}
    \delta_\text{H} = \frac{P_\text{H}(\text{primary}+\text{feed-down}) - P_\text{H}(\text{primary})}{P_\text{H}(\text{primary})},
    \label{FD-effect}
\end{equation}
from Eq.~(\ref{P-corr}). The feed-down contribution factors on the global $\Xi^-$ polarization in 20--50\% central Au+Au collisions at different energies are shown in Table~\ref{table-feed-down}. We find that the feed-down effect increases the measured global $\Xi^-$ polarization from the primary polarization by about 25\%. This value has a weak dependence on the collision energy due to the change of chemical freeze-out parameters~\cite{Adamczyk:2017iwn}.

By a similar method, we also calculate the feed-down effect on the global $\Lambda$ polarization. In the calculation, we consider $\Lambda(1405)$, $\Lambda(1520)$, $\Lambda(1600)$, $\Lambda(1670)$, $\Lambda(1690)$, $\Sigma^0$, $\Sigma(1385)$, $\Sigma(1660)$, $\Sigma(1670)$, $\Xi$, and $\Xi(1530)$ as the parent particles. Their primary yields are calculated by the THERMUS model. We assume that the primary global polarizations of all spin-1/2 particles are the same, and the primary global polarizations of spin-3/2 particles are $5/3$ times of that of spin-1/2 particles. The primary global polarizations of these particles are transferred to the decay product $\Lambda$'s through the rules listed in Table~\ref{table-C}.

The feed-down contribution factors on the global $\Lambda$ polarization in 20--50\% central Au+Au collisions at different energies are also shown in Table~\ref{table-feed-down}. We find that the feed-down effect can reduce the global $\Lambda$ polarization by 8--13\%. In the past studies~\cite{Becattini:2016gvu,Karpenko:2016jyx,Li:2017slc}, this decreasing was estimated to be 15--20\%, which is larger than our current estimation. The difference is mainly because that an important two-step decay $\Xi(1530)\to\Xi\to\Lambda$ was not considered in those studies.

\begin{figure}
    \centering
    \includegraphics[width=1\columnwidth]{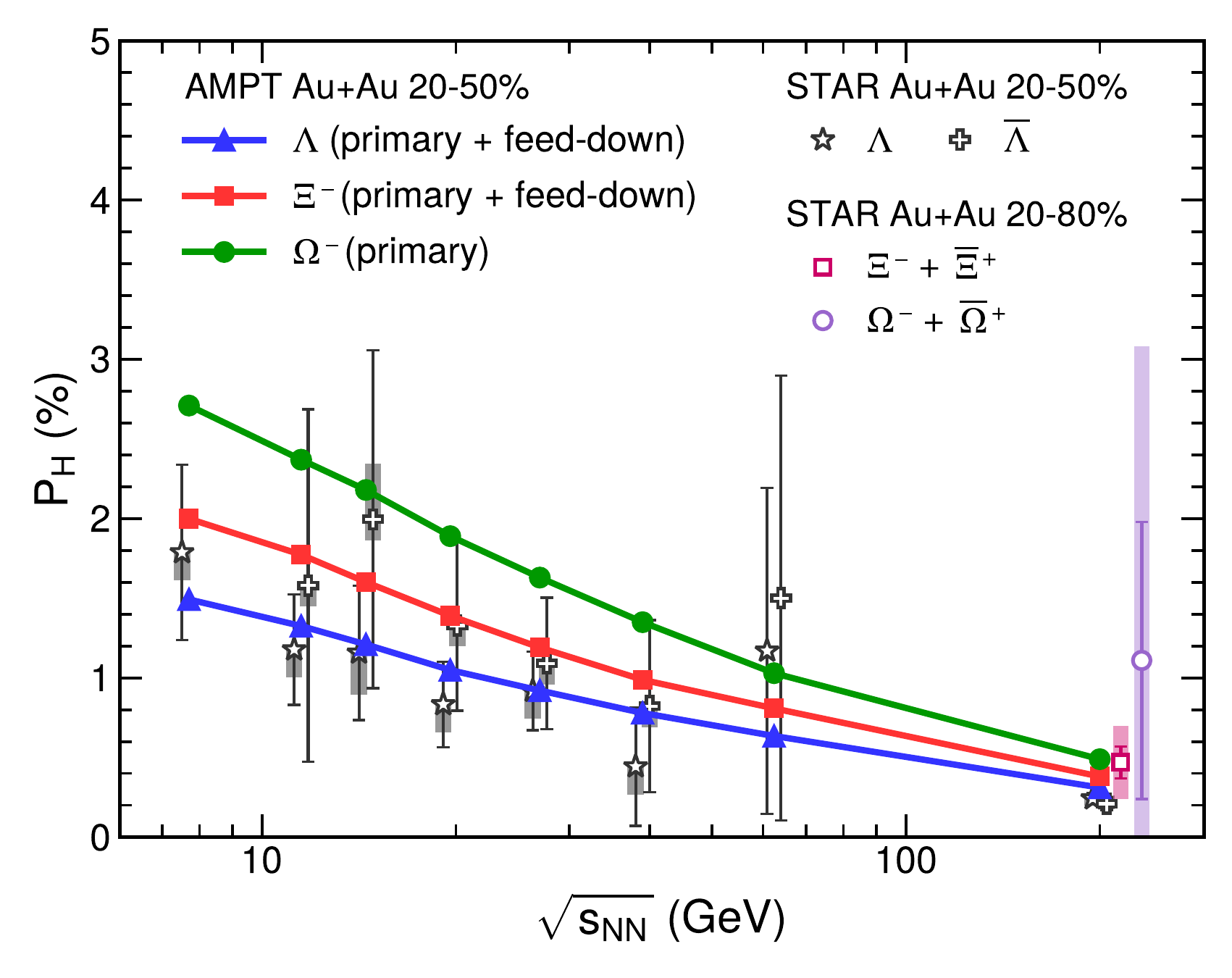}
    \caption{(Color online) Same as figure~\ref{fig-PH}, but the feed-down effect on the global polarizations of $\Lambda$ and $\Xi^-$ is taken into account.}
    \label{fig-FD}
\end{figure}

The global polarizations of $\Lambda$ and $\Xi^-$ with the feed-down contribution and the primary global polarization of $\Omega^-$ are shown in Fig.~\ref{fig-FD}. The global polarizations of $\Lambda$ and $\Xi^-$ are separated after the feed-down effect is taken into account. Finally, the global polarizations of three hyperons fulfill $P_{\Omega^-}>P_{\Xi^-}>P_\Lambda$. Such a global-polarization ordering is consistent with the experimental measurement at 200 GeV~\cite{Adam:2020pti} though the statistical errors are too large to be definitive. Future measurements from the RHIC beam energy scan program may shed new light on this finding.

\section{Conclusions}\label{sec4}

In summary, we have studied the global polarizations of $\Lambda$, $\Xi^-$, and $\Omega^-$ hyperons in Au+Au collisions at $\sNN=$ 7.7--200 GeV. Our calculations show that the primary global polarizations of $\Lambda$, $\Xi^-$, and $\Omega^-$ satisfy the approximate relation $P_{\Omega^-}\simeq 5/3P_{\Xi^-}\simeq 5/3P_\Lambda$, and after taking the feed-down effect into account, the global polarizations exhibit the ordering: $P_{\Omega^-}>P_{\Xi^-}>P_\Lambda$. In this ordering, $P_{\Omega^-}>P_{\Xi^-}$ and $P_{\Omega^-}>P_\Lambda$ are primarily due to the higher spin of $\Omega^-$ hyperon, while $P_{\Xi^-}>P_\Lambda$ is mainly because of the opposite feed-down effect on $\Lambda$ and $\Xi^-$. The global-polarization ordering found in this study is consistent with the recent experimental data at 200 GeV~\cite{Adam:2020pti}. More experimental measurements at other beam energies will further test our findings and provide insights into the global polarization mechanism.

\section*{Acknowledgments}

We thank T.~Niida for discussions. This work is supported by National Natural Science Foundation of China (NSFC) through grants No.~11835002 and No.~12075061 and Shanghai Natural Science Foundation through Grant No.~20ZR1404100. X.-L.~X.~and H.~L.~are also funded by China Postdoctoral Science Foundation through grants No.~2018M641909 and No.~2019M661333.

\bibliographystyle{elsarticle-num}
\bibliography{ref.bib}

\end{document}